# On the Evolution of Scalar Metric Perturbations in an Inflationary Cosmology

R. R. Caldwell

*University of Cambridge, D.A.M.T.P.*
*Silver Street, Cambridge CB3 9EW, U.K.*
*email: R.R.Caldwell@amtp.cam.ac.uk*

(September 15, 1995)

## Abstract

We further clarify how scalar metric perturbations are amplified in an inflationary cosmology. We first construct a simple, analytic model of an inflationary cosmology in which the expansion scale factor evolves continuously from an inflationary era to a radiation-dominated era. From this model, it is clear to see how scalar perturbations are amplified. Second, we examine the recent claims of Grishchuk, and the reply by Deruelle and Mukhanov, regarding the evolution of scalar perturbations through an abrupt transition in the equation of state of the cosmological fluid. We demonstrate that the "standard results" regarding the amplification of scalar, density perturbations from inflation are valid.

## I. INTRODUCTION

The goal of this paper is to understand the recent claims by Grishchuk [1–3] that the standard results on the amplification of scalar metric perturbations in an inflationary cosmology are incorrect. By standard result, we mean that the magnitude of the scalar perturbations is amplified by a factor determined by the change in the equation of state $w \equiv p/\rho$ of the cosmological fluid, where $p$ and $\rho$ are the background pressure and energy density. Crudely, the amplification is proportional to

$$\frac{1+w_+}{1+w_-} \qquad (1.1)$$

when the perturbations evolve through consecutive expansion eras in which the equation of state of the cosmological fluid is given by $w_-$, $w_+$. For $|1+w_-| \ll 1$, as occurs in an inflationary epoch, this amplification may be large.

It was argued in references [1–3] that the amplification of density perturbations was much weaker than commonly assumed. However, this claim is incorrect, as has been pointed out by Deruelle and Mukhanov [4]. There it was shown that the error in Grishchuk's argument was due to a subtlety in the junction conditions necessary to evolve the density perturbations across an abrupt transition in the equation of state of the cosmological fluid at the end of the inflationary epoch.

While the issue would seem to be settled, we found it useful to study a cosmological scenario which made no recourse to junction conditions. Hence, we constructed a toy cosmological model which evolves smoothly from an inflationary- to radiation-dominated expansion era. Although the evolution of density perturbations in this model must be carried out numerically, implementation of the code is a straight forward task. Hence, we confirmed that the standard results (as presented in [5] for example) were indeed valid. We hope that this paper will contribute towards the understanding of the mechanism by which scalar metric perturbations are amplified in an inflationary cosmology.

The contents of this paper are as follows. In section II, we construct a toy cosmological model and numerically evolve the scalar metric perturbations across the inflation-radiation transition. In section III, we point out the shortcomings in Grishchuk's argument, making reference to the work of Deruelle and Mukhanov. In section IV we turn to the cosmological significance of scalar metric perturbations. We demonstrate that scalar perturbations do not vanish in the de Sitter limit, contrary to Grishchuk's claim, but in fact are strongly amplified. We conclude in section V. Throughout this paper we use units in which the speed of light $c = 1$.

## II. TOY MODEL INFLATIONARY COSMOLOGY

In this section we construct an analytic, toy model inflationary cosmology. This model possesses the generic features of an inflationary cosmology; an arbitrarily long vacuum-dominated expansion era followed by a smooth transition to a radiation-dominated era, and strong amplification of scalar perturbations. We refer to this model as a "toy" because we have not made contact with a sensible particle physics model; this model is proposed for simple, demonstrative purposes.

### A. Cosmological Model

Consider a spatially-flat, isotropic FRW space-time with line element

$$ds^2 = -dt^2 + a^2(t)[dx^2 + dy^2 + dz^2]. \tag{2.1}$$

Here, $t$ is the physical time, and $a(t)$ is the expansion scale factor. The matter content of the space-time is given by

$$\rho(t) = \frac{3L^{-2}}{8\pi G} \frac{1}{a_1^4 + a^4(t)}. \tag{2.2}$$

Here, $a_1$ is a dimensionless constant, and $L$ is a length scale. Examining the above equation, we see that for small values of the scale factor, $a \ll a_1$, the energy density is approximately constant, as in a de Sitter, vacuum-dominated space-time. For large values of the scale-factor, $a \gg a_1$, the energy density decays as $a^{-4}$, as in a radiation-dominated space-time.

Thus, we have a "two-fluid" cosmology which evolves from vacuum- to radiation-domination with increasing $a$. We may adjust the energy density around the time of the transition from vacuum- to radiation-domination, by varying $a_1$. A simple extension of this toy model is to add another component which decays as $a^{-3}$, as in a dust-dominated space-time. In such a case, the energy density would be

$$\rho = \frac{3L^{-2}}{8\pi G}\Big(\frac{1}{a_1^4 + a^4} + \frac{a_2}{a_1^3 + a^3}\Big)$$

where $a_2$ is a dimensionless constant which determines the time of the transition from radiation- to dust-domination. For simplicity, we will only consider the toy model energy density given by equation (2.2).

The remaining details of the cosmology may be obtained from the Einstein equations. The Hubble rate is given by

$$H \equiv \frac{1}{a}\frac{da}{dt} = L^{-1}(a_1^4 + a^4)^{-1/2}. \tag{2.3}$$

The pressure of the cosmological fluid is

$$p = \frac{L^{-2}}{8\pi G}\frac{a^4 - 3a_1^4}{(a_1^4 + a^4)^2} = \frac{1}{3}\frac{a^4 - 3a_1^4}{a_1^4 + a^4}\rho. \tag{2.4}$$

From the equation for the energy density and pressure, we see that the inequality $\rho + 3p < 0$ is satisfied for $a < a_1$. Thus, the space-time will inflate for $a < a_1$. We are also interested in the quantities

$$w \equiv \frac{p}{\rho} = \frac{1}{3}\Big(\frac{a^4 - 3a_1^4}{a^4 + a_1^4}\Big)$$
$$\gamma \equiv \frac{dp}{d\rho}\Big|_{entropy} = \frac{dp/dt}{d\rho/dt} = -\frac{1}{3} + 2w. \tag{2.5}$$

Here, $w$ gives the equation of state of the cosmological fluid, and $\gamma$ is the adiabatic index.

An advantage of this toy model, for instructional purposes, is that it makes no reference to a scalar field or potential which may compose the vacuum-dominated cosmological fluid. As well, all quantities are expressed in terms of the expansion scale factor $a$ and the dimensionless constant $a_1$. There is no need to obtain the time dependence of the scale factor.

### B. Adiabatic Density Perturbations

We consider scalar perturbations to the spatially flat FRW metric [5]

$$ds^2 = a^2(\eta)[-d\eta^2(1 + 2\phi) + 2B_{;i}dx^i d\eta + dx^i dx^j(2E_{;ij} + (1 - 2\psi)\delta_{ij})], \tag{2.6}$$

where $\eta$ is the conformal time. In the above line element, the derivatives are with respect to the spatially flat background 3−metric. We may construct combinations of the scalar quantities $\phi$, $\psi$, $E$, $B$ which are invariant under a general coordinate transformation $x^\alpha \to \tilde{x}^\alpha = x^\alpha + \xi^\alpha$. These invariant combinations

$$\Phi = \phi + \frac{1}{a}[(B - E')a']'$$

$$\Psi = \psi - \frac{a'}{a}(B - E') \tag{2.7}$$

help simplify the evolution equations for density perturbations. Here $' \equiv \partial/\partial\eta$. Similarly, we may consider the perturbed Einstein equations under a general coordinate transformation and construct gauge invariant quantities. These equations may be written in terms of $\Phi$ and $\Psi$, and gauge invariant combinations of the source stress-energy tensor. For a symmetric source stress-energy tensor, a further simplification occurs, as the field equations require $\Phi = \Psi$. The quantity $\Phi = \Phi(\eta, \vec{x})$ is referred to as the gauge invariant potential, and characterizes the amplitude of density perturbations. On scales below the Hubble radius, $\Phi$ plays a role similar to the Newtonian potential.

The evolution of the gauge invariant potential for the case of adiabatic perturbations is given by the equation

$$\frac{d^2\Phi}{d\eta^2} + 3\mathcal{H}(1+\gamma)\frac{d\Phi}{d\eta} - \gamma\nabla^2\Phi + [2\frac{d\mathcal{H}}{d\eta} + (1+3\gamma)\mathcal{H}^2]\Phi = 0. \tag{2.8}$$

By adiabatic perturbations, we mean perturbations to the cosmological fluid $\delta\rho$ and $\delta p$ such that $\delta p = \gamma\delta\rho$. In the above evolution equation, $\eta$ is the conformal expansion time, and $\mathcal{H} \equiv d\ln a/d\eta$. A clear derivation of this evolution equation is found in the review by Mukhanov, Feldman, and Brandenberger [5]. We would like to follow the evolution of $\Phi$ from the vacuum- to radiation-dominated era.

We will recast the evolution equation using the expressions $dt = ad\eta$ and $da/dt = aH$ to relate the conformal and physical time to the scale factor. We obtain

$$\frac{d^2\Phi}{da^2} + \frac{1}{a}(\frac{7}{2} + 3\gamma - \frac{3}{2}w)\frac{d\Phi}{da} - \frac{\gamma}{a^4 H^2}\nabla^2\Phi + \frac{1}{a^2}3(\gamma - w)\Phi = 0. \tag{2.9}$$

Thus, we have eliminated the need for a time coordinate, expressing the evolution in terms of the expansion scale factor. Note that this form of the evolution equation is generally valid; we have not used any results specific to our toy cosmology in arriving at (2.9).

We next decompose the gravitational potential into spatial Fourier harmonics.

$$\Phi(a, \vec{x}) = \int d^3k \tilde{\Phi}(a, \vec{k}) e^{i\vec{k}\cdot\vec{x}} \tag{2.10}$$

Thus we replace $\Phi \to \tilde{\Phi}$ and $\nabla^2 \to -k^2$ in equation (2.9). This decomposition will allow us to follow the evolution of a single mode through the vacuum- and radiation-dominated expansion eras.

Let us qualitatively examine the evolution equation (2.9). At early and late times, we may use equations (2.5) to simplify the form of the evolution equation.

$$\text{For} \quad a \ll a_1, \quad \frac{d^2\tilde{\Phi}}{da^2} - \frac{2}{a}\frac{d\tilde{\Phi}}{da} - (\frac{7a_1^4}{3a^4}k^2 L^2 + \frac{4}{a^2})\tilde{\Phi} = 0$$

$$\text{For} \quad a \gg a_1, \qquad \frac{d^2\tilde{\Phi}}{da^2} + \frac{4}{a}\frac{d\tilde{\Phi}}{da} + (\frac{1}{3}k^2 L^2 - \frac{4a_1^4}{a^6})\tilde{\Phi} = 0 \qquad (2.11)$$

At early times, $\gamma < w$, and $\tilde{\Phi}$ has a rapidly growing mode. Physically, the quickly changing negative vacuum pressure of the space-time couples to the gauge invariant potential, serving as a strong anti-damping or "pump" source. Hence, we expect rapid growth in $\tilde{\Phi}$ at early times. This is the mechanism by which the scalar perturbations are amplified. The equation of state is $|1+w| \approx \frac{4}{3}(a/a_1)^4 \ll 1$, indicating strong amplification according to the expression (1.1). At late times, $\gamma \sim w$, and the coefficient of the term linear in $\tilde{\Phi}$ is negligibly small for $k^2 L^2 \ll a_1^4/a^6$, indicating a constant solution. Hence, we expect the rapid growth to cease and the amplitude of $\tilde{\Phi}$ to stabilize at late times. While this is indeed what happens, we are not interested in a crude, qualitative estimate. We proceed to the next section for an exact solution to the evolution.

### C. Numerical Solution

The evolution of the gauge invariant potential as a function of the expansion scale factor may easily be obtained numerically. It is a straight forward exercise to convert equation (2.9) into a pair of first order differential equations. Specifying the initial data $(\tilde{\Phi}, d\ln\tilde{\Phi}/d\ln a)$ at some initial time $a_i$, we may integrate to obtain $\tilde{\Phi}$ at a later time $a_f$. The results are presented in figure 1. Curves showing the evolution of $\tilde{\Phi}$ are presented in figure 1, for initial values of $d\ln\tilde{\Phi}/d\ln a = (0, 1, 10)$, with $k \ll H$, representing scalar fluctuations with wavelengths longer than the Hubble length. The potential grows rapidly, as $\tilde{\Phi} \propto a^4$ in the vacuum-dominated era. However, upon the onset of the radiation-dominated era, for $a \gtrsim a_1$, the growth stops and $\tilde{\Phi}$ remains constant. The evolution is just as we expected from our qualitative discussion above. Hence, we have constructed a toy cosmological model in which the magnitude of scalar perturbations is strongly amplified.

## III. STANDARD TECHNIQUES

In this section we examine the recent claims of Grishchuk [1–3], and the reply by Deruelle and Mukhanov [4], regarding the evolution of scalar perturbations across a phase transition. We demonstrate that the "standard results" regarding the amplification of scalar, density perturbations from inflation are valid.

### A. $\zeta$ as a Conserved Quantity

A standard technique for estimating the amplitude of the gauge invariant potential $\tilde{\Phi}$ is to use the integral of the equation of motion, $\zeta$. As has been well demonstrated elsewhere [5], the quantity

$$\zeta \equiv \frac{2}{3}\frac{\tilde{\Phi} + \mathcal{H}^{-1}\tilde{\Phi}'}{1+w} + \tilde{\Phi} \qquad (3.1)$$

is approximately constant in time for modes $k \ll H$. Thus, one may use $\zeta$ as a sort of conserved quantity. Applying this technique to the case of our toy model cosmology, we obtain

$$\zeta(a) = \zeta(a_i) \longrightarrow \tilde{\Phi} = \tilde{\Phi}_i \Big(\frac{a}{a_i}\Big)^4 \Big(\frac{3a_i^4 + a_1^4}{3a^4 + a_1^4} + \frac{a_i^4 + a_1^4}{3a^4 + a_1^4}\frac{d\ln\tilde{\Phi}_i}{d\ln a_i}\Big). \tag{3.2}$$

Here, we have assumed that $d\ln\tilde{\Phi}/d\ln a \ll 1$. A comparison of the evolution of $\tilde{\Phi}$ using equation (3.2) with the exact, numerical result is presented in figure 2. We see that the use of $\zeta$ as a conserved quantity to obtain the evolution of $\tilde{\Phi}$ is well justified.

It has been claimed by Grishchuk [2] that $\zeta = 0$ for all times, and therefore conveys no information as a conserved quantity. We will examine this argument closely. In equation (38′) of [2], the evolution equation for the synchronous gauge scalar quantity $\mu$ is given as

$$\frac{1}{a^2\gamma_G}\Big[a^2\gamma_G\Big(\frac{\mu}{a\sqrt{\gamma_G}}\Big)'\Big]' + n^2\frac{\mu}{a\sqrt{\gamma_G}} = X. \tag{3.3}$$

Here $\gamma_G = 1 - \mathcal{H}'/\mathcal{H}^2$, $n$ is the mode number (referred to in these notes as $k$), and

$$\mu = -\frac{1}{2}\frac{a}{\mathcal{H}\sqrt{\gamma_G}}(\psi' + \mathcal{H}\gamma_G\psi) \tag{3.4}$$

with $\psi$ defined by equation (2.6). Next, taking the definition of $\zeta$ given by [5], expressed in terms of synchronous gauge variables, Grishchuk writes

$$\zeta_{MFB} = \frac{1}{2n^2}\frac{1}{a^2\gamma_G}\Big[a^2\gamma_G\Big(\frac{\mu}{a\sqrt{\gamma_G}}\Big)'\Big]'. \tag{3.5}$$

The argument proceeds as follows. Take the long wavelength limit, for which $\zeta$ is claimed to be conserved, letting $n \to 0$ in equation (3.3). Insert the resulting expression for $X$ into the equation defining $\zeta$, to obtain

$$\zeta_{MFB} = \frac{X}{2n^2}. \tag{3.6}$$

Note that (3.3) is simply the equation of motion for the scalar perturbations, with $X = 0$. Hence, Grishchuk claims, $\zeta_{MFB} = 0$. However, the error in this argument is apparent. The $n \to 0$ limit has not been taken consistently. Returning to (3.3), using $X = 0$, we substitute the expression for $n^2\mu/a\sqrt{\gamma_G}$ into (3.5) to obtain

$$\zeta_{MFB} = -\frac{\mu}{2a\sqrt{\gamma_G}}. \tag{3.7}$$

Relating $\mu$ to our more familiar gauge invariant potential $\Phi$, we find that this equation is equivalent to (3.1). Hence, this exercise has simply used the equations of motion to recast the form of $\zeta$.

The quantity $\zeta$ may be zero for specially chosen data. For a given background spacetime, specifying the matter content or the expansion scale factor, one may always choose a set of initial data $(\tilde{\Phi}_i, d\ln\tilde{\Phi}_i/d\ln a_i)$ satisfying the equation of motion (2.9), for which $\zeta = 0$. However, the equations of motion do not demand $\zeta = 0$.

## B. Matching Conditions for the Evolution Across an Abrupt Transition

In the recent work of Deruelle and Mukhanov [4], the junction conditions were obtained which are required to patch together the evolution of the gauge invariant potential across a phase transition in which the equation of state of the cosmological fluid changes discontinuously. The scenario considered is the following. The gauge invariant potential evolves according to equation (2.9), until the adiabatic index $\gamma$ and the equation of state $w$ change discontinuously, after which the evolution is again described by (2.9). To match the solutions $\tilde{\Phi}_-$ and $\tilde{\Phi}_+$ (the $\pm$ refers the value on either side of the transition) across the transition, they required the continuity of the total energy and momentum of the cosmological fluid. Defining the transition to occur when the total pressure, the background plus perturbations, reaches a critical value, they showed that the surface of transition is not in general a surface of constant time. Transforming into a coordinate system in which the transition surface occurs at constant time, they obtain the matching conditions, in terms of synchronous gauge quantities $\psi$ and $E$:

$$[\psi + \mathcal{H}E']_\pm = 0 \qquad [\zeta - \frac{2}{9\mathcal{H}^2(1+w)}\nabla^2(\psi + \mathcal{H}E')]_\pm = 0. \tag{3.8}$$

In the synchronous gauge, chosen here to make contact with Grishchuk's work, $\phi = B = 0$ in equations (2.6) and (2.7)). We see that the matching conditions reproduce the result that $\zeta$ serves as a conserved quantity for long wavelength perturbations, and is constant across the transition.

As an application of these matching conditions, consider a power-law expansion epoch, such that $\gamma = w$. Given initial conditions ($\tilde{\Phi}_i$, $d\ln\tilde{\Phi}_i/d\ln a_i$), the evolution for $\tilde{\Phi}$ is

$$\tilde{\Phi} \approx \tilde{\Phi}_i\Big(1 + \frac{2}{5+3\gamma}\frac{d\ln\tilde{\Phi}_i}{d\ln a_i}\Big[1 - \big(\frac{a_i}{a}\big)^{\frac{1}{2}(5+3\gamma)}\Big]\Big) \tag{3.9}$$

for long wavelength modes. Employing the matching conditions (3.8) to continue the evolution into another power-law expansion epoch we find

$$\frac{d\ln\tilde{\Phi}_+}{d\ln a} = \Big(\frac{w_+ - w_-}{w_- + 1}\Big) + \Big(\frac{w_+ + 1}{w_- + 1}\Big)\frac{d\ln\tilde{\Phi}_-}{d\ln a}. \tag{3.10}$$

For $|1 + w_-| \ll 1$, as occurs at the end of an inflationary epoch, the derivative $d\tilde{\Phi}/da$ term will receive a substantial amplification. The evolution of $\tilde{\Phi}$ through vacuum- ($w = -1 + \delta$, $0 < \delta \ll 1$) and radiation-dominated ($w = 1/3$) expansion epochs is displayed in figure 3. While no substantial growth occurs for the long wavelength modes during either epoch, the abrupt transition leads to a strong amplification. The late-time amplification obtained by modeling the transition from vacuum- to radiation-dominated expansion by a discontinuous jump in the equation of state of the cosmological fluid agrees with the exact result obtained for the cosmological model used in section II.

Grishchuk has proposed a different set of matching conditions for the evolution of the gauge invariant potential across an abrupt transition. In terms of Grishchuk's quantities, Deruelle and Mukhanov's matching conditions (3.8) are written as

$$[-h + n^{-2}\mathcal{H}h_l'Q]_{\pm} = 0 \qquad [\zeta - \frac{4}{9\mathcal{H}^2(1+w)}\nabla^2(-h + n^{-2}\mathcal{H}h_l'Q)]_{\pm} = 0, \qquad (3.11)$$

Where $Q(\vec{x})$ are the scalar harmonics on $R^3$. In terms of our notation, $h = -\psi/2$ and $h_l = n^2 E/2$. However, Grishchuk [1] has matched the scalar quantities $h$, $h_l$ on a surface of constant time which happens not to be a surface of constant energy. This amounts to requiring that the scalar quantities $h$, $h_l$ and their time derivatives match across the transition:

$$[h]_{\pm} = 0 \qquad [h']_{\pm} = 0 \qquad [h_l]_{\pm} = 0 \qquad [h_l']_{\pm} = 0. \qquad (3.12)$$

These conditions are clearly inequivalent to those obtained by Deruelle and Mukhanov, which return the standard result.

An additional set of matching conditions are discussed by Grishchuk. Following equation (48) of [1], the evolution of the scalar perturbations across the abrupt transition requires

$$[\mu]_{\pm} = 0 \qquad [v]_{\pm} = 0 \qquad (3.13)$$

as the matching conditions. The quantity $\mu$ has been defined in equation (3.4), and

$$v = -n^2 h + h_l' = \frac{n^2}{2}(\mathcal{H}^{-1}\psi + E'). \qquad (3.14)$$

We see that matching $v$ is consistent with (3.12). Next, using (3.7) for $\zeta$ in terms of $\mu$, and (3.4) for $\mu$ in terms of $h$, $h'$, we rewrite the additional matching condition as

$$[\mu]_{\pm} = [-2a\sqrt{\gamma_G}\zeta]_{\pm} = [\frac{a}{\mathcal{H}\sqrt{\gamma_G}}(h' + \mathcal{H}\gamma_G h)]_{\pm} = 0. \qquad (3.15)$$

It is clear from the middle term above that this condition is inequivalent to the correct condition (3.8), since $\zeta$ is conserved for long wavelengths, whereas $\gamma_G$ changes discontinuously across the abrupt transition. From the term on the right above, we see that matching $\mu$ is inequivalent to matching $h$, $h'$. Hence, the conditions (3.13) are inequivalent to the conditions actually employed by Grishchuk, (3.12), and the correct conditions (3.8).

The evolution of $\tilde{\Phi}$ through consecutive power-law expansion epochs, using both conditions (3.12) and (3.13), are shown in figure 3. Clearly, these choices of matching conditions are incorrect, as they disagree with both the exact results, and those results obtained through other techniques.

We take this opportunity to comment briefly on the matching conditions for cosmological perturbations at a transition in the equation of state prescribed by Ratra [7]. The condition given for the evolution of the density perturbation $\delta\rho$, equations (3.10) and (4.8) in [7] agree with the results of Deruelle and Mukhanov [4] only in the longitudinal gauge. The results do not agree, and are incorrect when applied in the synchronous gauge. The reason is that the density perturbation itself is a gauge dependent quantity; Ratra's condition for the joining of $\delta\rho$ is not gauge invariant.

## IV. STANDARD RESULTS

In this brief section we make contact with the standard results on the amplitude of perturbations produced by inflation. We first discuss how the amplitude of scalar perturbations is determined. Second, we consider the amplitude of perturbations in the limiting case of a de Sitter inflationary era.

### A. Observable Quantities

The physical significance of the gauge invariant potential may be understood if we consider the temperature anisotropy induced in the cosmic microwave background (CMB). On large scales, the dominant source of the temperature anisotropy due to the gravitational potential at the surface of last scattering [6] is

$$\frac{\delta T}{T}(\hat{\Omega}) \approx \frac{1}{3}\Phi(\eta_{\text{ls}}, |\eta_{\text{o}} - \eta_{\text{ls}}|\hat{\Omega}). \tag{4.1}$$

Thus, the CMB anisotropy pattern is determined by the gauge invariant potential $\Phi$. In this equation the potential is evaluated at the time of last scattering $\eta_{\text{ls}}$, where the present time is $\eta_{\text{o}}$, $\eta_{\text{o}} - \eta_{\text{ls}}$ is the conformal distance to the surface of last scattering, and $\hat{\Omega}$ is a unit vector on the sphere.

From our experience in the previous sections, we may use the quantity $\zeta$ to evolve the potential $\Phi$ from the inflationary to matter-dominated era. Hence, for long wavelengths, $\Phi_m$ in the matter era is given by

$$\Phi_m \approx \frac{2}{5}\frac{\Phi_i + \mathcal{H}^{-1}\Phi'_i}{1 + w_i} \tag{4.2}$$

where $w_i$, $\Phi_i$ are evaluated in the inflationary era. Next we must set the initial amplitude of the potential.

In a realistic model of inflation, scalar metric perturbations are generated according to the following crude scheme. The vacuum energy of a homogeneous scalar field $\phi_0$ drives the inflationary expansion; quantum mechanical fluctuations $\delta\varphi$ around the background scalar field produce scalar metric fluctuations $\Phi$ [5,6]; the scalar perturbations are amplified by the vacuum-dominated expansion. As we have already examined the subsequent evolution of the scalar perturbations, we need to fix the initial conditions for the potential $\Phi$. We refer to Einstein's equations to express the fluctuations around the background source in terms of metric perturbations (see [5]),

$$\mathcal{H}\Phi + \Phi' = 4\pi G\phi'_0\delta\varphi \qquad \rightarrow \qquad \Phi_m \approx \frac{3}{5}\mathcal{H}\frac{\delta\varphi}{\phi'_0}|_i \tag{4.3}$$

where we have used the relationship $1 + w = (8\pi G/3)\mathcal{H}^{-2}\phi'^2_0$ for the scalar field-dominated equation of state. The amplitude of $\delta\varphi$ is determined by the equal time, two-point function $\langle\delta\varphi(x)\delta\varphi(x')\rangle$ in the appropriate quantum mechanical vacuum state. Hence, the amplitude of the CMB anisotropy, and the gauge invariant potential, is determined by the amplitude of the quantum mechanical fluctuations in the background scalar field. Equation (4.3) for $\Phi_m$ agrees with the standard result for the amplitude of perturbations from inflation [5,8].

## B. The Limit of de Sitter Inflation

We now examine the case of a de Sitter inflationary era, taken as the limit of a general power-law expansion era. Consider a power-law expansion scale factor $a(\eta) \propto \eta^{-1-\delta}$ where $\eta$ is the conformal time, as $\delta \to 0$. In this well-studied scenario [6], the time variation of the background scalar field

$$\phi_0' \propto \delta^{1/2} \qquad (4.4)$$

vanishes in the de Sitter, $\delta \to 0$ limit. Returning to (4.3), the term which fixes the amplitude of the potential during the de Sitter era, the product $\phi_0' \delta\varphi$, appears to vanish. However, let us also examine the rms amplitude of $\delta\varphi$. In this scenario,

$$\langle \delta\varphi^2 \rangle = \frac{\eta}{8\pi a^2(\eta)} \int_0^\infty dk\, k^2 |H^{(1)}_{\frac{3}{2}-\delta}(k\eta)|^2. \qquad (4.5)$$

In the de Sitter limit, it is well known that the rms amplitude diverges logarithmically, due to the behavior of the integrand at small $k$. Examining the integral more closely, we find that in the limit $\delta \to 0$, the product $\phi_0' \langle \delta\varphi^2 \rangle^{1/2}$ does not vanish. Hence, the term in (4.3) which fixes the amplitude of the potential does not vanish in the de Sitter limit.

We may now return to the claims by Grishchuk [1–3] that the magnitude of scalar perturbations vanishes in the limit of a de Sitter inflationary era. The fault in the argument is that the divergence in the rms amplitude of the fluctuations in the scalar field has been ignored. As we have demonstrated above, the amplitude of scalar perturbations is in fact finite during the de Sitter epoch. However, these scalar perturbations experience infinite amplification at the close of the inflationary era, since the factor $1 + w_-$ in (1.1) vanishes. Note that the calculation of the evolution of scalar perturbations has been carried out within the context of linearized gravity, under the assumption that all perturbations are small, $|\Phi| \ll 1$. Hence, there is a limit to how closely the inflationary era may consistently approximate a de Sitter space-time. Generically, this restriction is satisfied by inflationary cosmologies for which the amplitude of the induced temperature anisotropies is consistent with observations.

## V. SUMMARY

We have confirmed the validity of the standard results on the evolution of scalar perturbations in an inflationary cosmology. Our main result is the demonstration that the gauge invariant potential $\Phi$ is strongly amplified during the course of evolution from a vacuum- to radiation-dominated expansion era. Next, we have demonstrated that $\zeta$, the integral of the time dependent portion of the equation of motion, may be used to excellent approximation as a tool for computing the evolution of long wavelength modes of $\Phi$.

We have also discussed the recent claims of Grishchuk [1–3] that scalar perturbations are not strongly amplified in an inflationary cosmology. First, we demonstrated that the claim $\zeta = 0$ for all times is incorrect. Second, making reference to the work of Deruelle and Mukhanov [4], we have demonstrated the flaw in Grishchuk's matching conditions for the evolution of the gauge invariant potential across an abrupt transition. The standard result is consistent with the conservation of $\zeta$ for long wavelength perturbations. Third, we have shown that scalar perturbations do not vanish in the limit of a de Sitter inflationary epoch. Rather, the perturbations are strongly amplified as the limit is approached, as given in the standard literature [4–6].

## ACKNOWLEDGMENTS

We would like to thank Bruce Allen, Nathalie Deruelle, Juan Garcia-Bellido, Leonid Grishchuk, Andrew Liddle, Dave Salopek, John Stewart, and David Wands for useful conversations during the course of this investigation. The work of RRC was supported by PPARC through grant number GR/H71550.

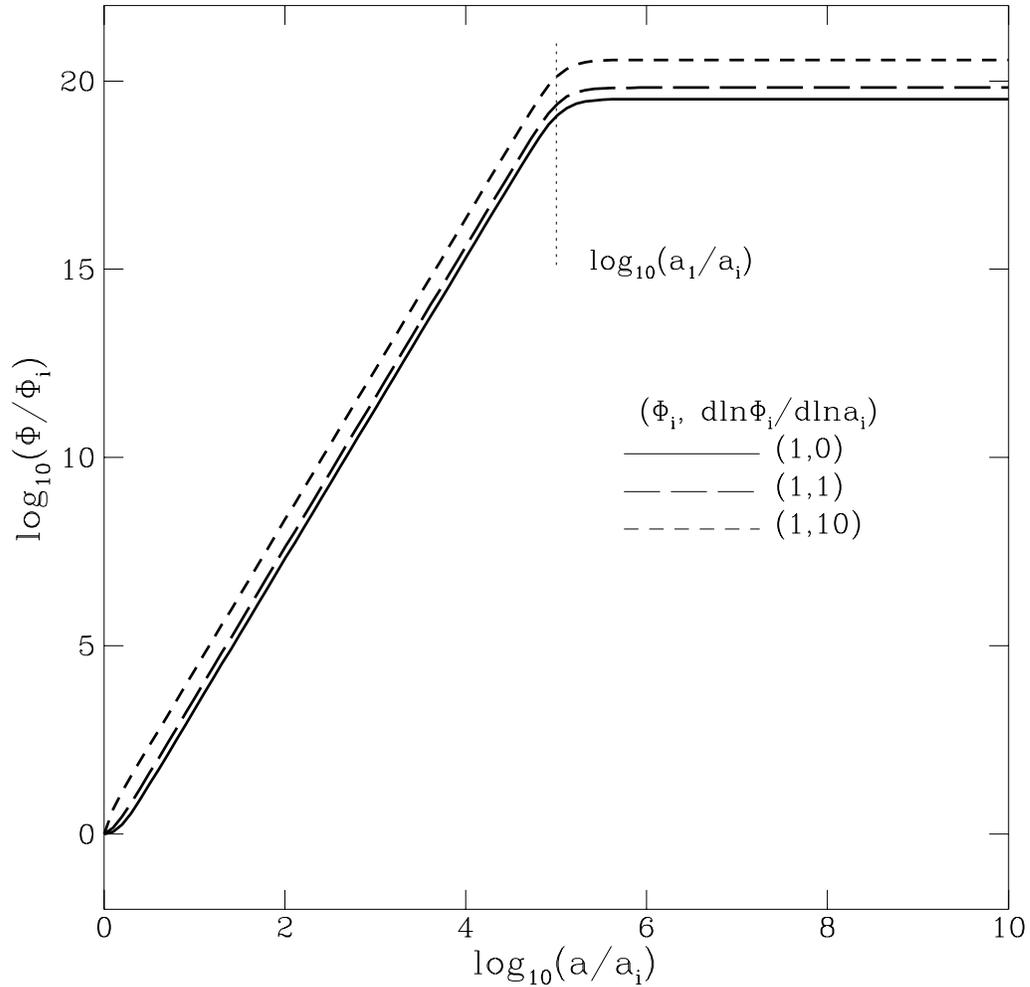

FIG. 1. The evolution of the gauge invariant potential $\Phi$ in the toy cosmological model constructed in these notes is displayed. The evolution is computed exactly, for three sets of initial conditions, for long wavelength modes. The expansion is vacuum-dominated for $a < a_1$, and radiation-dominated for $a_1 < a$.

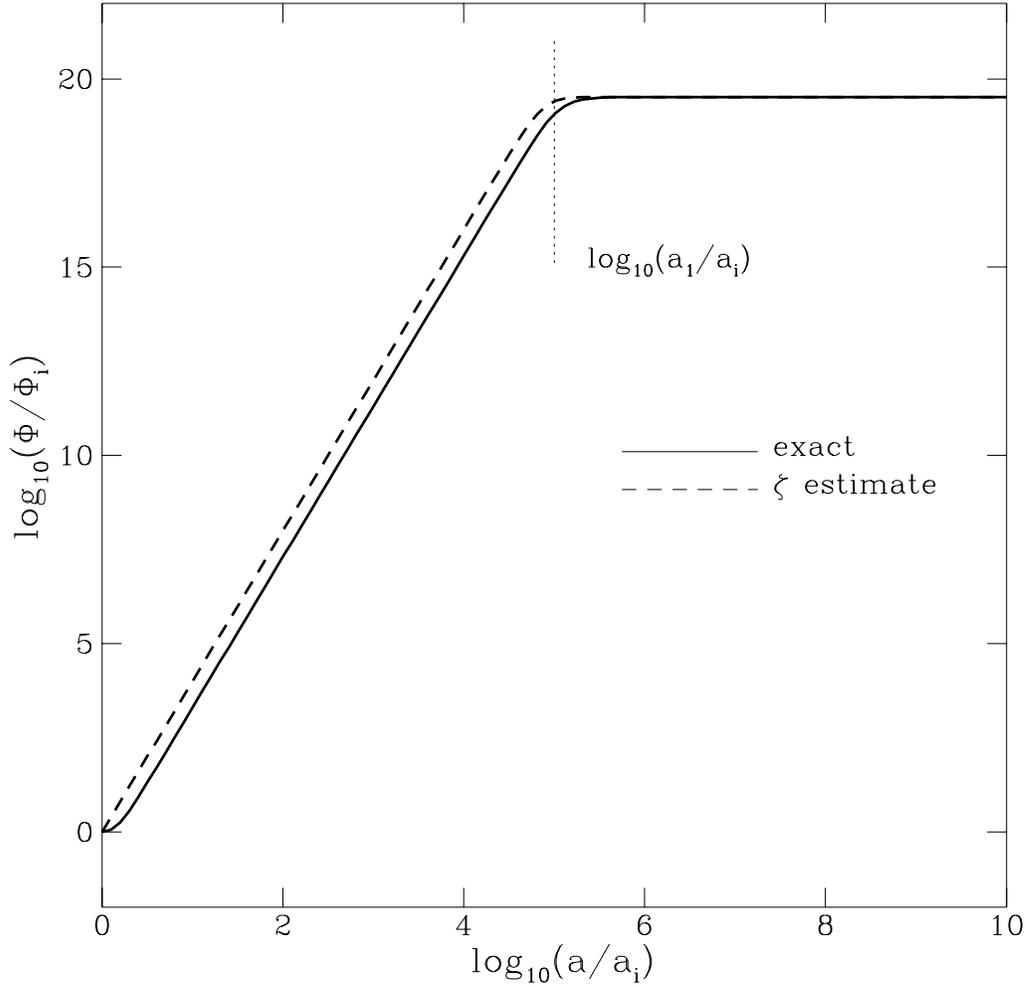

FIG. 2. The evolution of the gauge invariant potential $\Phi$ in the toy cosmological model constructed in these notes is displayed. The solid curve shows the exact evolution. The dashed curve shows the evolution obtained using $\zeta$ as a conserved quantity. The disagreement for $a < a_1$ occurs because we have ignored the contribution of the derivative term $d\ln\Phi/d\ln a$ in $\zeta$. The agreement is excellent at late times, when $|d\ln\Phi/d\ln a| \ll 1$ and the derivative term is negligible.

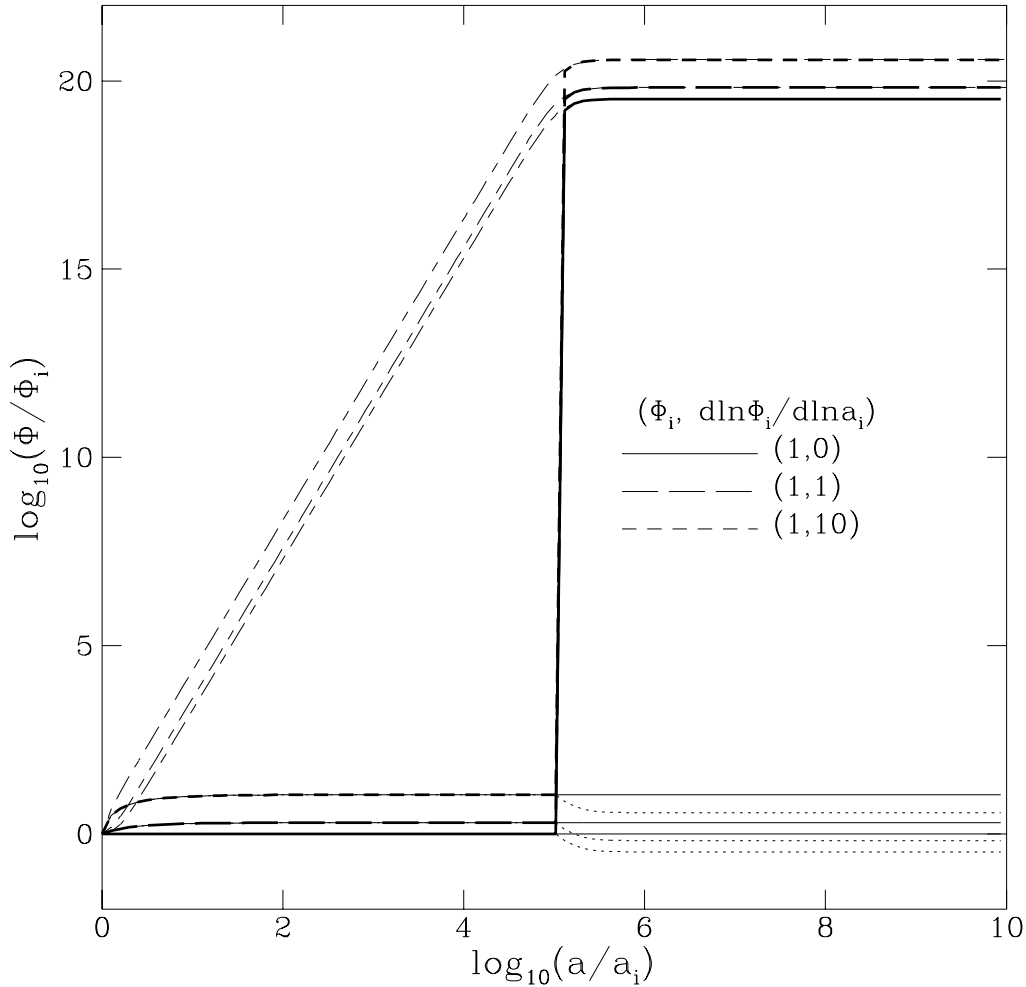

FIG. 3. The evolution of the gauge invariant potential $\Phi$ in a cosmological model constructed from consecutive, power-law expansion epochs, matching the potential across the abrupt transition. The evolution is computed exactly, for three sets of initial conditions, for long wavelength modes. The matching conditions obtained by Deruelle and Mukhanov, using $\zeta$ as a conserved quantity, were used to evolve $\Phi$ across the transition. The light, solid lines show the evolution of $\Phi$ using the matching conditions (3.12) used by Grishchuk. The light, dotted lines show the evolution of $\Phi$ using the matching conditions (3.13) discussed by Grishchuk. The light, long-short dashed lines show the exact evolution of $\Phi$ from figure 1. The expansion is vacuum-dominated for $a < a_1$ ($w = -1 + \frac{4}{3}(a_i/a_1)^4$), and radiation-dominated for $a_1 < a$ ($w = 1/3$). Similar results were obtained for a toy model which evolves from vacuum- to dust-dominated expansion.